\documentclass[pra, sd, amsmath,amssymb, preprint]{revtex4-1}

\usepackage{txfonts}
\usepackage{hyperref}
\usepackage{graphicx}
\usepackage{dcolumn}
\usepackage{bm}
\usepackage{float}
\usepackage{color}
\usepackage{braket}

\draft 

\begin{document}


\title{Finite-Temperature Many-Body Perturbation Theory in the Canonical Ensemble} 

\author{Punit K. Jha}%
\author{So Hirata}%
  \email{sohirata@illinois.edu}
  \affiliation{ 
Department of Chemistry, University of Illinois at Urbana-Champaign, Urbana, Illinois 61801, USA
}




\date{\today}

\begin{abstract}
Benchmark data are presented for the zeroth- through third-order many-body perturbation corrections to the 
electronic Helmholtz energy, internal energy, and entropy in the canonical ensemble 
 in a wide range of temperature. They are determined  as numerical $\lambda$-derivatives 
of the respective quantities computed by thermal full configuration interaction with a perturbation-scaled Hamiltonian,
$\hat{H}=\hat{H}_0+\lambda\hat{V}$. Sum-over-states analytical formulas for up to the third-order corrections to these properties
are also derived as analytical $\lambda$-derivatives. These formulas, which are verified by exact numerical agreement with the benchmark data, are given in terms of 
the Hirschfelder--Certain degenerate perturbation energies and should be valid for both degenerate and nondegenerate reference states at any temperature down to zero. The 
results in the canonical ensemble are compared with the same in the grand canonical ensemble.
\end{abstract}


\maketitle 


\section{Introduction} 

In a previous study \cite{jha}, we reported the benchmark data for several low-order perturbation corrections to the electronic grand potential, internal energy, and chemical potential
 of an ideal gas of molecules in the grand canonical ensemble. They were determined numerically as the $\lambda$-derivatives of the respective quantities calculated exactly, i.e., by thermal full configuration interaction (FCI) \cite{Kou2014} with a perturbation-scaled Hamiltonian, $\hat{H}_0+\lambda\hat{V}$. We call this the $\lambda$-variation method \cite{Hirata2017}. The first- and second-order corrections evaluated by the finite-temperature many-body perturbation theoretical formulas in a number of textbooks \cite{thouless1972quantum,bla86,mattuck1992guide,march1995many,fetter2003quantum,SANTRA} 
were shown to disagree with these benchmark data, implying that the theory is incorrect beyond the zeroth order
and does not converge at the exact limit. This failure was ascribed not so much to any mathematical issue as to its neglect
of the variation of chemical potential with $\lambda$, causing the average number of electrons to fluctuate 
and violating the net electrical neutrality of the system as a basic tenet of equilibrium thermodynamics \cite{Fisher,Dyson,HirataARPC,Levin}. 

We derived \cite{so_jha} the correct first-order correction formulas for the grand potential, internal energy, and chemical potential  in the grand canonical ensemble by
demanding to restore the electrical neutrality of the system at each perturbation order.
These analytical formulas were given in two forms:\ sum-over-states expressions written in 
terms of the energy corrections according to the Hirschfelder--Certain degenerate perturbation theory (HCPT) \cite{hc_dpt} and reduced formulas expressed with molecular integrals
and the Fermi--Dirac distribution function.
They both reproduce the benchmark data at any temperature down to zero.
The latter were derived from the former using the sum rules of the HCPT corrections and several Boltzmann-sum identities, one of which being responsible for the same kind of 
massive mathematical simplifications in the zeroth-order (Fermi--Dirac) theory. These simplifications use nothing more than elementary calculus and combinatorics as well as the HCPT sum rules
and there is no need to resort to the Matsubara Green's function \cite{matsubara} or thermal Wick's theorem \cite{sanyal} in the time-dependent diagrammatic logic, which seems much less tractable. 

Another way to restore the electrical neutrality of the system is simply to adopt the canonical ensemble and sample only the electrically neutral states. 
There have been only a few attempts to formulate finite-temperature many-body perturbation theory in the canonical ensemble of electrons \cite{infinite_series,fermi_canonical,deviations}. Arnaud {\it et al.}\ \cite{infinite_series}
and Sch\"{o}nhammer \cite{fermi_canonical} derived the exact thermodynamic properties of noninteracting electrons by assuming equidistant energy levels. 
However, there have been no reports of order-by-order analytic equations or numerical values of the perturbation corrections to various thermodynamic quantities in 
the canonical ensemble in a more general case. 
This is probably because the kind of elegant mathematical simplifications in the grand canonical ensemble, which leads to 
the Fermi--Dirac theory at the zeroth order or similar analytical formulas at the first order \cite{so_jha}, does not seem to occur \cite{deviations} in the canonical ensemble. 

In this work, we use the $\lambda$-variation method \cite{Hirata2017} to determine the benchmark numerical data for the zeroth- through 
third-order perturbation corrections to the Helmholtz energy, internal energy, and entropy in the canonical ensemble of electrons in ideal gases of identical atoms or molecules. We then present sum-over-states analytical formulas for the zeroth- through third-order perturbation corrections to these thermodynamic quantities, given in terms of the HCPT energy 
expressions \cite{hc_dpt}. These formulas are, again, obtained as analytical $\lambda$-derivatives of thermal FCI expressions in a purely time-independent, nondiagrammatic derivation.
We show that they exactly reproduce the benchmark numerical data and thus form the basis of finite-temperature perturbation theory in the canonical ensemble.
We also make a comparison of the perturbation corrections of these properties between the canonical and grand canonical ensembles. 

\section{Thermal full configuration interaction} 

Let us consider an ideal gas of neutral atoms or molecules in the canonical ensemble.
Its electronic partition function $Z$ (ignoring the vibrational, rotational, and translational partition functions) at temperature $T$ is defined by
\begin{eqnarray} 
Z =   \sum_{I=1}^{{}_{2k}C_N} e^{ -\beta E_I },
\label{eq:C_PF_used}
\end{eqnarray} 
where $\beta = (k_\text{B} T)^{-1} $ and $E_I$ is the exact (i.e., zero-temperature-FCI) energy of the $I$th state. 
The sum is taken over all ${}_{2k}C_N$ states with $N$ electrons (where $N$ is the sum of the atomic numbers in the molecule) occupying $2k$ spinorbitals spanned by a set of $k$ basis functions.
The Helmholtz energy $F$, internal energy $U$, and entropy $S$ are related to $Z$ by
\begin{eqnarray}
F &=& - \frac{1}{\beta} \ln Z,
\label{eq:C_F} \\
U &=& - \frac{\partial}{ \partial \beta} \ln Z,
\label{eq:C_U} 
\end{eqnarray}
and
\begin{eqnarray}
S &=& - k_\text{B} \sum_{I=1}^{{}_{2n}C_N} {W_I} \ln {W_I}
\label{eq:C_S}
\end{eqnarray}
with 
\begin{eqnarray}
{W_I} = \frac{e^{-\beta E_I}}{Z}.
\label{eq:C_W}
\end{eqnarray}
They are also related to one another by the identity,
\begin{eqnarray}
S = k_\text{B} \beta( U - F ) .
\label{eq:C_S2}
\end{eqnarray}
We call this computational procedure to determine the values of 
$F$, $U$, and $S$ the thermal FCI method \cite{Kou2014}, which constitutes the numerically exact electronic thermodynamics of an ideal gas within a basis set.

\section{$\lambda$-variation numerical benchmarks} 

The $n$th-order correction $X^{(n)}$ of thermodynamic property $X$ ($X=F$, $U$, or $S$ in this case) 
is defined \cite{Hirata2017} as the $n$th derivative with respect to $\lambda$ of the same property $X(\lambda)$ determined exactly by thermal FCI using 
a perturbation-scaled Hamiltonian $\hat{H}=\hat{H}_0 + \lambda\hat{V}$,
\begin{eqnarray}
X^{(n)} =\left. \frac{1}{n!}\frac{\partial^{n}X(\lambda)}{\partial \lambda^{n}}  \right\vert_{\lambda=0},
\label{eq:C_lambda_F}
\end{eqnarray}
where $\hat{H}_0$ is the zeroth-order Hamiltonian, $\hat{V}$ is the perturbation, and $\lambda=1$ corresponds to the fully interacting system.  
This exactly matches with the usual perturbation expansion of $X(\lambda)$,
\begin{eqnarray}
X(\lambda) = X^{(0)} + \lambda X^{(1)} + \lambda^2 X^{(2)} + \dots, 
\end{eqnarray}
forming a converging series toward thermal FCI at $\lambda = 1$. A finite-difference evaluation of $X^{(n)}$ for several low orders
consists in the $\lambda$-variation method \cite{Hirata2017}, furnishing invaluable numerical benchmark data for any perturbation theory with any partitioning
of Hamiltonian or any reference wave function. There is a minimal risk of programming or formulation errors with this method.

We applied this method to ideal gases of the hydrogen fluoride molecule (0.9164 \AA, $N=10, k=6$), the boron hydride molecule (1.232 \AA, $N=6, k=6$), and the beryllium atom ($N=4, k=5$) 
in the minimal (STO-3G) basis set. 
We adopted the M{\o}ller--Plesset partitioning of the Hamiltonian, where $\hat{H}_0$ is the zero-temperature Fock operator plus the nuclear-repulsion energy.  Hence, 
the reference wave function was the zero-temperature $N$-electron ground-state Hartree--Fock wave function. Throughout the calculations, molecular orbitals and orbital energies were held fixed. 

We used the seven-point central finite-difference formula \cite{Fornberg} at $\lambda=0$ with the grid spacing of $\Delta\lambda = 10^{-2}$ for the first and second derivatives (yielding the first- and second-order
perturbation corrections) and $\Delta\lambda = 10^{-1}$ for the third derivatives (furnishing the third-order perturbation corrections). 
These parameter choices were made on the basis of an analysis of the results using a wide range of their values. 
Equations (\ref{eq:C_S}) and (\ref{eq:C_W}) were 
evaluated when computing $S^{(n)}$, which were cross-checked against $U^{(n)}$ and $F^{(n)}$ using Eq.\ (\ref{eq:C_S2}).  

The zeroth- through third-order perturbation corrections to the Helmholtz energy $F$, internal energy $U$, and entropy $S$ of the ideal gas
of hydrogen fluoride in the canonical ensemble are documented in Tables \ref{tab:HF_F}, \ref{tab:HF_U}, and \ref{tab:HF_S}, respectively. They are also compared with 
the perturbation corrections to grand potential $\Omega$, internal energy $U$, and entropy $S$ of the identical system in the grand canonical ensemble \cite{jha}.
Table \ref{tab:diff} shows the convergence of the perturbation series
towards thermal FCI \cite{Kou2014}. The results are discussed in Sec.\ \ref{Discussion}.

The data for the boron hydride and beryllium can be found in the Appendix. 

\begin{table*} 
\caption{ \label{tab:HF_F} The zeroth- through third-order perturbation corrections to the Helmholtz energy $F$ as a function of temperature $T$ 
obtained from the $\lambda$-variation method for an ideal gas of hydrogen fluoride in the canonical ensemble as well as the zeroth- through second-order perturbation corrections
to grand potential $\Omega$ in the grand canonical ensemble. }
\begin{ruledtabular}
\begin{tabular}{lddddddd}
& \multicolumn{4}{c}{Canonical ensemble} & \multicolumn{3}{c}{Grand canonical ensemble\tablenotemark[1]} \\ \cline{2-5} \cline{6-8} 
$T /~\text{K}$ &  \multicolumn{1}{c}{ $F^{(0)}/E_\text{h}$ } &  \multicolumn{1}{c}{ $F^{(1)}/E_\text{h}$} & \multicolumn{1}{c}{$F^{(2)}/E_\text{h}$} & \multicolumn{1}{c}{$F^{(3)}/E_\text{h}$}  
& \multicolumn{1}{c}{ $\Omega^{(0)}/E_\text{h}$ } &  \multicolumn{1}{c}{ $\Omega^{(1)}/E_\text{h}$} & \multicolumn{1}{c}{$\Omega^{(2)}/E_\text{h}$} \\ \hline
$10^3$ & -52.5749 & -45.9959 & -0.0173 & -0.0055& -53.4112 & -45.9959& -0.4353 \\
$10^4$ & -52.5749 & -45.9959 & -0.0173 & -0.0055& -53.5117 & -45.9959& -0.4324 \\
$10^5$ & -52.6717 & -46.1631 & -0.1466 & -0.0524& -55.6365 & -45.2684& -2.5815 \\
$10^6$ & -62.5554 & -46.7786 & -0.0165 &  0.0003& -105.947 & -44.5256& -0.9643 \\
$10^7$ & -176.803 & -46.8574 & -0.0024 &  0.0000& -686.703 & -43.1991& -0.1970 \\
$10^8$ & -1368.93 & -46.8576 & -0.0004 &  0.0000& -6804.94 & -41.9847& -0.0276 \\
$10^9$ & -13309.7 & -46.8555 & -0.0000 &  0.0000& -68084.5 & -41.8264& -0.0029 
\end{tabular}
\tablenotetext[1]{Reference \onlinecite{jha}.}
\end{ruledtabular}
\end{table*}

\begin{table*}  
\caption{ \label{tab:HF_U} The same as Table \ref{tab:HF_F} but for the internal energy $U$. }
\begin{ruledtabular}
\begin{tabular}{lddddddd}
& \multicolumn{4}{c}{Canonical ensemble} & \multicolumn{3}{c}{Grand canonical ensemble\tablenotemark[1]} \\ \cline{2-5} \cline{6-8} 
$T /~\text{K}$ &  \multicolumn{1}{c}{ $U^{(0)}/E_\text{h}$ } &  \multicolumn{1}{c}{ $U^{(1)}/E_\text{h}$} & \multicolumn{1}{c}{$U^{(2)}/E_\text{h}$} & \multicolumn{1}{c}{$U^{(3)}/E_\text{h}$}  
& \multicolumn{1}{c}{ $U^{(0)}/E_\text{h}$ } &  \multicolumn{1}{c}{ $U^{(1)}/E_\text{h}$} & \multicolumn{1}{c}{$U^{(2)}/E_\text{h}$} \\ \hline
$10^3$ & -52.5749 & -45.9959 & -0.0173 & -0.0055& -52.5749 & -45.9959 & -0.0173 \\
$10^4$ & -52.5749 & -45.9959 & -0.0173 & -0.0055& -52.5749 & -45.9959 & -0.0173 \\
$10^5$ & -52.2645 & -45.6944 & -0.0215 & -0.1665& -52.0166 & -45.9479 &  0.0984 \\
$10^6$ & -50.6228 & -46.7166 & -0.0342 &  0.0009& -50.5964 & -46.1767 & -0.2198 \\
$10^7$ & -46.0028 & -46.8452 & -0.0037 &  0.0001& -45.7891 & -46.2355 & -0.0326 \\
$10^8$ & -42.4046 & -46.8596 & -0.0008 &  0.0000& -42.3641 & -46.1180 & -0.0054 \\
$10^9$ & -41.9496 & -46.8557 & -0.0001 &  0.0000& -41.9453 & -46.0975 & -0.0006
\end{tabular}
\tablenotetext[1]{Reference \onlinecite{jha}.}
\end{ruledtabular}
\end{table*}

\begin{table*} 
\caption{ \label{tab:HF_S} The same as Table \ref{tab:HF_F} but for the entropy $S$. }
\begin{ruledtabular}
\begin{tabular}{lddddddd}
& \multicolumn{4}{c}{Canonical ensemble} & \multicolumn{3}{c}{Grand canonical ensemble\tablenotemark[1]} \\ \cline{2-5} \cline{6-8} 
$T /~\text{K}$ &  \multicolumn{1}{c}{ $S^{(0)}/k_\text{B}$ } &  \multicolumn{1}{c}{ $S^{(1)}/k_\text{B}$} & \multicolumn{1}{c}{$S^{(2)}/k_\text{B}$} & \multicolumn{1}{c}{$S^{(3)}/k_\text{B}$}  
& \multicolumn{1}{c}{ $S^{(0)}/k_\text{B}$ } &  \multicolumn{1}{c}{ $S^{(1)}/k_\text{B}$} & \multicolumn{1}{c}{$S^{(2)}/k_\text{B}$} \\ \hline
$10^3$ & 0.0000 & 0.0000  & 0.0000  & 0.0000  & 0.0000 & 0.0000 & 0.0000 \\
$10^4$ & 0.0000 & 0.0000  & 0.0000  & 0.0000  & 0.0000 & 0.0000 & 0.0000 \\
$10^5$ & 1.2856 & 1.4801  & 0.3949  & -0.3602 & 2.8344 & 0.2288 & 1.1370 \\
$10^6$ & 3.7680 & 0.0196  & -0.0056 & 0.0002  & 4.9697 & 0.0122 &-0.0336 \\
$10^7$ & 4.1304 & 0.0004  & 0.0000 & 0.0000  & 5.3498 &-0.0018 &-0.0004 \\
$10^8$ & 4.1889 & 0.0000 & 0.0000  & 0.0000  & 5.4060 &-0.0000 &-0.0000 \\
$10^9$ & 4.1896 & 0.0000  & 0.0000  & 0.0000  & 5.4067 & 0.0000 & 0.0000 
\end{tabular}
\tablenotetext[1]{Reference \onlinecite{jha}.}
\end{ruledtabular}
\end{table*}

\begin{table}  
\caption{ \label{tab:diff} The difference of the sum of zeroth- through third-order corrections  
from the thermal-FCI value \cite{Kou2014} for the Helmholtz energy $F$, internal energy $U$, or entropy $S$
as a function of temperature $T$ for an ideal gas of hydrogen fluoride in the canonical ensemble. }
\begin{ruledtabular}
\begin{tabular}{lddd}
$T /~\text{K}$ &  \multicolumn{1}{c}{$  \Delta F /E_\text{h}$} &  \multicolumn{1}{c}{$ \Delta U /E_\text{h}$} & \multicolumn{1}{c}{$ \Delta S/k_\text{B}$}   \\ \hline
$10^3$ & 0.0030  & 0.0030 & 0.0000  \\
$10^4$ & 0.0030  & 0.0030 & -0.0001 \\
$10^5$ & -0.0133 & 0.0314 & 0.1414  \\
$10^6$ & 0.0000  & 0.0000 & 0.0000  \\
$10^7$ & 0.0000  & 0.0000 & 0.0000  \\
$10^8$ & 0.0000  & 0.0000 & 0.0000  \\
$10^9$ & 0.0000  & 0.0000 & 0.0000  
\end{tabular}
\end{ruledtabular}
\end{table}

\section{Sum-over-states analytical formulas}

In this section, sum-over-states analytical formulas for the zeroth- through third-order perturbation corrections to $F$, $U$, and $S$ are presented.
Tables \ref{tab:C_F} and \ref{tab:C_U} compare the perturbation corrections to $F$ and $U$, respectively, 
calculated by these analytical formulas with the $\lambda$-variation benchmark data.

These analytical formulas are derived by analytical $\lambda$-differentiation of the exact (thermal-FCI) expressions of the respective quantities 
as per Eq.\ (\ref{eq:C_lambda_F}). They are expressed in terms of the perturbation corrections to the FCI energies,
$\{E_I^{(n)}\}$, where $E_I^{(n)}$ is the $n$th-order correction to the zeroth-order energy of the $I$th state according to HCPT \cite{hc_dpt}; we 
cannot rely on M{\o}ller--Plesset perturbation theory (MPPT) \cite{shavitt2009many} because many excited states are exactly degenerate
at the zeroth order. 
For nondegenerate states, HCPT reduces to MPPT if the  M{\o}ller--Plesset partitioning of the Hamiltonian
is adopted. In either case, these energy corrections conform to the canonical definition of perturbation corrections as given by Eq.\ (\ref{eq:C_lambda_F}) with $X = E_I$.

We have not found the kind of drastic simplification (as in the Fermi--Dirac theory) which brings these sum-over-states formulas involving long
sums over exponentially many states into more compact ones involving much shorter sums over molecular integrals and 
the Fermi--Dirac distribution function (as in the finite-temperature perturbation theory
in the grand canonical ensemble \cite{so_jha}). However, they may still serve in practice at low temperatures, where 
these long sums may be  truncated aggressively with minimal errors.

We use the following two Taylor-series expressions \cite{so_jha},
\begin{eqnarray}
e^{a+b}&=& e^{a} + be^a + \frac{b^2}{2!} e^a + \frac{b^3}{3!} e^a + \dots, \label{eq:exp}\\
\ln(a+b)&=& \ln a+ \frac{b}{a} - \frac{b^2}{2a^2} + \frac{b^3}{3a^3}+  \dots,
\label{eq:ln}
\end{eqnarray}
which are rapidly convergent when $a \gg b$ . 

\subsection{Zeroth order}
The zeroth-order canonical partition function is given by 
\begin{eqnarray}
Z^{(0)}  =\sum_I e^{-\beta E_I^{(0)}},
\label{eq:zeroth_one}
\end{eqnarray}
where $E_I^{(0)}$ is the zeroth-order HCPT energy of the $I$th state. It can also be written as
\begin{eqnarray}
E_I^{(0)} = E_{\text{nuc.}}+\sum_{i}^{\text{occ.}} \epsilon_i,
\end{eqnarray}
where $E_{\text{nuc.}}$ is the nuclear-repulsion energy, $\epsilon_i$ is the $i$th orbital energy, and the summation runs over all orbitals occupied by an electron in the $I$th state. 

Then, according to Eqs.\ (\ref{eq:C_F}) and (\ref{eq:C_U}), we have
\begin{eqnarray}
F^{(0)}  &=&  -\frac{1}{\beta} \ln Z^{(0)} = -\frac{1}{\beta} \ln \sum_I  e^{-\beta E_I^{(0)}}, 
\label{eq:zeroth_three}\\
U^{(0)}  &=&  -\frac{\partial}{\partial \beta} \ln Z^{(0)} =  \frac{\sum_I E_I^{(0)} e^{-\beta E_I^{(0)}}}{\sum_I e^{-\beta E_I^{(0)}}} = \left \langle E^{(0)} \right \rangle,
\label{eq:zeroth_four}
\end{eqnarray}
where $\langle \dots \rangle$ denotes a canonical ensemble average,
\begin{eqnarray}
\Big \langle X \Big \rangle \equiv \frac{\sum_I X_I e^{-\beta E_I^{(0)}}}{\sum_I e^{-\beta E_I^{(0)}}}.
\end{eqnarray}

The zeroth-order entropy is given by 
\begin{eqnarray}
S^{(0)} = k_\text{B}\beta \left(U^{(0)} - F^{(0)} \right),
\end{eqnarray}
according to Eq.\ (\ref{eq:C_S2}).

\subsection{First order}

Applying Eq.\ (\ref{eq:exp}) to Eq.\ (\ref{eq:C_PF_used}) and collecting terms that are first order in $\lambda$, we find
\begin{eqnarray}
Z^{(1)}  = \sum_I \left( -\beta E_I^{(1)} \right) e^{-\beta E_I^{(0)}},
\label{eq:first_one}
\end{eqnarray}
where $E_I^{(1)}$ is the first-order HCPT energy correction \cite{hc_dpt} of the $I$th state.
Using Eq.\ (\ref{eq:ln}), we can write the first-order corrections to the Helmholtz and internal energies as well as entropy as
\begin{eqnarray}
F^{(1)}  &=&  -\frac{1}{\beta}\frac{Z^{(1)}}{Z^{(0)}}= \frac{\sum_I E_I^{(1)} e^{-\beta E_I^{(0)}}}{\sum_I e^{-\beta E_I^{(0)}}} = 
\left \langle E^{(1)}  \right \rangle
\label{eq:first_three}, \\
U^{(1)}  &=&  -\frac{\partial}{\partial \beta}  \left( \frac{Z^{(1)}}{Z^{(0)}} \right) 
= -\frac{\partial}{\partial \beta}  \left( -\beta F^{(1)} \right) 
\nonumber\\
&=& 
\left \langle E^{(1)}  \right \rangle + \beta \left \langle E^{(1)}  \right \rangle \left \langle E^{(0)}  \right \rangle 
-\beta\left \langle E^{(1)} E^{(0)} \right \rangle, 
\label{eq:first_four}
\end{eqnarray}
and
\begin{eqnarray}
S^{(1)} &=& k_\text{B}\beta \left(U^{(1)} - F^{(1)} \right) \nonumber\\
&=& k_\text{B}\beta^2 \left \langle E^{(1)}  \right \rangle \left \langle E^{(0)}  \right \rangle 
- k_\text{B}\beta^2\left \langle E^{(1)} E^{(0)} \right \rangle,
\end{eqnarray}
where we used
\begin{eqnarray}
\frac{\partial }{\partial \beta} \Big\langle X  \Big\rangle = \Big\langle X  \Big\rangle \left \langle E^{(0)} \right \rangle - \left \langle X  E^{(0)} \right \rangle. \label{X}
\end{eqnarray}

The last two terms of Eq.\ (\ref{eq:first_four}) individually scale quadratically with molecular size and are non-size-consistent. 
It is expected (albeit not proven) that these non-size-consistent contributions cancel exactly across the two terms, leaving only the size-consistent
contribution. That this is the case is implied by Eq.\ (\ref{X}) because the left-hand side is size-consistent, provided that the HCPT energy corrections are size-consistent.

Each of the terms that has at least one factor of $\beta$ multiplying $\langle \dots \rangle$ vanishes in the high-temperature ($\beta \to 0$) limit.
On the other hand, in the low- and high-temperature limits, we have
\begin{eqnarray}
\lim_{T \to 0} \Big\langle X \Big\rangle &=& X_0, \\
\lim_{T \to \infty} \Big\langle X \Big\rangle &=& \bar X \equiv 
 \frac{\sum_I X_I}{{}_{2k}C_{N}} , 
\end{eqnarray}
where $X_0$ is the value of $X$ for the $N$-electron ground (``zeroth'') state, and $\bar X$ is an unweighted average. 
Therefore,
\begin{eqnarray}
\lim_{T \to 0} F^{(n)} &=& \lim_{T \to 0} U^{(n)} = E_0^{(n)}, \label{F1} \\
\lim_{T \to \infty} F^{(n)} &=& \lim_{T \to \infty} U^{(n)} = \bar E^{(n)}, \\
\lim_{T \to 0} S^{(n)} &=& \lim_{T \to \infty} S^{(n)} = 0.\label{S1} 
\end{eqnarray}
where $n=1$. In fact, they hold for $n=2$ and $3$ (see below) and likely for all $n \geq 1$. 

\subsection{Second order}

Expanding Eq.\ (\ref{eq:C_PF_used}) into the form of Eq.\ (\ref{eq:exp}) and collecting terms that are second order in $\lambda$, we obtain
\begin{eqnarray}
Z^{(2)} &=& \sum_I \left( -\beta E_I^{(2)} + \frac{\beta^2}{2} E_I^{(1)}E_I^{(1)}  \right) e^{-\beta E_I^{(0)}}, \label{eq:sec_one}
\end{eqnarray}
where $E_I^{(2)}$ is the second-order HCPT energy correction \cite{hc_dpt} of the $I$th state. 
Using Eq.\ (\ref{eq:ln}), we find
\begin{eqnarray}
F^{(2)}  &=&  -\frac{1}{\beta}\frac{Z^{(2)}}{Z^{(0)}}+\frac{1}{2\beta} \left(   \frac{Z^{(1)}}{Z^{(0)}} \right)  ^2 \nonumber \\
       &=& \left \langle E^{(2)}  \right \rangle - \frac{\beta}{2} \left \langle E^{(1)} E^{(1)} \right \rangle
       +\frac{\beta}{2} \left \langle E^{(1)} \right \rangle \left \langle E^{(1)} \right \rangle,
\label{eq:sec_three}
\end{eqnarray}
and
\begin{eqnarray}
U^{(2)}  &=& -\frac{\partial}{\partial \beta}  \left( -\beta F^{(2)} \right) 
= F^{(2)} +\beta\,\frac{\partial F^{(2)}}{\partial \beta} \nonumber \\
  &=& 
\left \langle E^{(2)} \right \rangle - \beta \left \langle E^{(1)} E^{(1)} \right \rangle 
+ \beta \left \langle E^{(1)} \right \rangle \left \langle E^{(1)} \right \rangle
       \nonumber\\&&
       +\beta \left \langle E^{(2)}  \right \rangle \left \langle E^{(0)}  \right \rangle 
- \beta \left \langle E^{(2)} E^{(0)} \right \rangle
       \nonumber\\&&
       - \frac{\beta^2}{2} \left \langle E^{(1)} E^{(1)} \right \rangle  \left \langle E^{(0)} \right \rangle
              + \frac{\beta^2}{2} \left \langle E^{(1)} E^{(1)} E^{(0)} \right \rangle      
               \nonumber\\&&
              + \beta^2 \left \langle E^{(1)} \right \rangle \left \langle E^{(1)} \right \rangle \left \langle E^{(0)} \right \rangle
              - \beta^2 \left \langle E^{(1)} \right \rangle \left \langle E^{(1)} E^{(0)} \right \rangle,
\label{eq:sec_four}
\end{eqnarray}
which also utilized Eq.\ (\ref{X}). The second-order entropy correction then reads
\begin{eqnarray}
S^{(2)}  &=& 
k_\text{B}\beta \left(U^{(2)} - F^{(2)} \right) \nonumber \\
&=& - \frac{k_\text{B}\beta^2}{2} \left \langle E^{(1)} E^{(1)} \right \rangle 
+ \frac{k_\text{B}\beta^2}{2}  \left \langle E^{(1)} \right \rangle \left \langle E^{(1)} \right \rangle
       \nonumber\\&&
       +k_\text{B}\beta^2 \left \langle E^{(2)}  \right \rangle \left \langle E^{(0)}  \right \rangle 
-k_\text{B} \beta^2 \left \langle E^{(2)} E^{(0)} \right \rangle
       \nonumber\\&&
       - \frac{k_\text{B}\beta^3}{2} \left \langle E^{(1)} E^{(1)} \right \rangle  \left \langle E^{(0)} \right \rangle
              + \frac{k_\text{B}\beta^3}{2} \left \langle E^{(1)} E^{(1)} E^{(0)} \right \rangle      
               \nonumber\\&&
              +k_\text{B} \beta^3 \left \langle E^{(1)} \right \rangle \left \langle E^{(1)} \right \rangle \left \langle E^{(0)} \right \rangle
               \nonumber\\&&
              - k_\text{B}\beta^3 \left \langle E^{(1)} \right \rangle \left \langle E^{(1)} E^{(0)} \right \rangle,
\label{eq:sec_five}
\end{eqnarray}
according to Eq.\ (\ref{eq:C_S2}).

Again, those terms containing a factor of $\beta$ or $\beta^2$ individually violate size-consistency, but
the non-size-consistent contributions are expected to cancel one another exactly, leaving a size-consistent remainder.
In fact, the foregoing expressions are reminiscent of 
Brueckner's bracket notation \cite{shavitt2009many,brueckner,goldstone},
which was used to prove the diagrammatic linkedness and thus size-consistency of zero-temperature MPPT.

The high- and low-temperature limits of the second-order corrections are given by the same equations (\ref{F1})--(\ref{S1}) with $n=2$.
It will furthermore be shown that $\bar{E}^{(2)}=0$ (see Sec.\ \ref{Discussion}).

\subsection{Third order}
Following the same procedure, we obtain the third-order correction to the canonical partition function, which reads
\begin{eqnarray}
Z^{(3)} 
= \sum_I  \left( -\beta E_I^{(3)} + \beta^2 E_I^{(1)}E_I^{(2)} - \frac{\beta^3}{3!} E_I^{(1)} E_I^{(1)}E_I^{(1)}  \right) e^{-\beta E_I^{(0)}}.
\nonumber\\ \label{eq:three_one}
\end{eqnarray}
The third-order corrections to the Helmholtz and internal energies are given by
\begin{eqnarray}
F^{(3)}  &=&  -\frac{1}{\beta}\frac{Z^{(3)}}{Z^{(0)}}  + \frac{1}{ \beta}\frac{Z^{(1)}}{Z^{(0)}}\frac{Z^{(2)}}{Z^{(0)}} - \frac{1}{3  \beta} \left( \frac{Z^{(1)}}{Z^{(0)}} \right)^3\nonumber \\
&=&  
\left \langle E^{(3)} \right \rangle - \beta \left \langle E^{(1)} E^{(2)} \right \rangle 
+ \frac{\beta^2}{3!} \left \langle E^{(1)}  E^{(1)}  E^{(1)} \right \rangle
\nonumber\\&&
+ \beta \left \langle E^{(1)}  \right \rangle \left \langle E^{(2)} \right \rangle 
-\frac{\beta^2}{2} \left \langle E^{(1)}  \right \rangle \left \langle E^{(1)}  E^{(1)} \right \rangle
\nonumber\\&&
+ \frac{\beta^2}{3}  \left \langle E^{(1)}  \right \rangle \left \langle E^{(1)}  \right \rangle \left \langle E^{(1)}  \right \rangle ,
\label{eq:three_three}
\end{eqnarray}
and
\begin{eqnarray}
U^{(3)}  &=& 
-\frac{\partial}{\partial \beta}  \left( -\beta F^{(3)} \right) 
= F^{(3)} +\beta\,\frac{\partial F^{(3)}}{\partial \beta}\nonumber \\
&=& 
\left \langle E^{(3)} \right \rangle 
- 2\beta \left \langle E^{(1)} E^{(2)} \right \rangle 
+\frac{\beta^2}{2} \left \langle E^{(1)}  E^{(1)}  E^{(1)} \right \rangle 
\nonumber\\&& 
+ 2 \beta \left \langle E^{(1)} \right \rangle  \left \langle E^{(2)} \right \rangle 
- \frac{3\beta^2}{2} \left \langle E^{(1)}  E^{(1)} \right \rangle \left \langle E^{(1)} \right \rangle
\nonumber\\&& 
+ \beta^2 \left \langle E^{(1)} \right \rangle \left \langle E^{(1)} \right \rangle \left \langle E^{(1)} \right \rangle 
+ \beta \left \langle E^{(3)} \right \rangle \left \langle E^{(0)} \right \rangle 
\nonumber\\&& 
- \beta \left \langle E^{(3)} E^{(0)} \right \rangle
- \beta^2 \left \langle E^{(1)}E^{(2)} \right \rangle \left \langle E^{(0)} \right \rangle 
+ \beta^2 \left \langle E^{(1)}E^{(2)} E^{(0)} \right \rangle
\nonumber\\&& 
+\frac{\beta^3}{3!} \left \langle E^{(1)}E^{(1)}E^{(1)} \right \rangle \left \langle E^{(0)} \right \rangle
-\frac{\beta^3}{3!} \left \langle E^{(1)}E^{(1)}E^{(1)} E^{(0)} \right \rangle
\nonumber\\&& 
+2\beta^2 \left \langle E^{(1)} \right \rangle  \left \langle E^{(2)} \right \rangle \left \langle E^{(0)} \right \rangle
- \beta^2 \left \langle E^{(1)} \right \rangle  \left \langle E^{(2)} E^{(0)} \right \rangle
\nonumber\\&& 
 - \beta^2 \left \langle E^{(1)}  E^{(0)} \right \rangle \left \langle E^{(2)} \right \rangle
- {\beta^3} \left \langle E^{(1)}  E^{(1)} \right \rangle \left \langle E^{(1)} \right \rangle \left \langle E^{(0)} \right \rangle
\nonumber\\&& 
+ \frac{\beta^3}{2} \left \langle E^{(1)}  E^{(1)} \right \rangle \left \langle E^{(1)} E^{(0)} \right \rangle
  + \frac{\beta^3}{2} \left \langle E^{(1)}  E^{(1)} E^{(0)} \right \rangle \left \langle E^{(1)} \right \rangle 
\nonumber\\&& 
+ \beta^3 \left \langle E^{(1)} \right \rangle \left \langle E^{(1)} \right \rangle \left \langle E^{(1)} \right \rangle \left \langle E^{(0)} \right \rangle 
\nonumber\\&& 
- \beta^3 \left \langle E^{(1)} \right \rangle \left \langle E^{(1)} \right \rangle \left \langle E^{(1)}  E^{(0)} \right \rangle .
\label{eq:three_four}
\end{eqnarray}
We will not give the lengthy expanded expression of $S^{(3)}$ here because it is easily reproduced from 
\begin{eqnarray}
S^{(3)} = k_\text{B}\beta \left(U^{(3)} - F^{(3)} \right).
\end{eqnarray}

The non-size-consistent contributions (the terms multiplied by a power of $\beta$)
are, again, expected to mutually cancel one another. 
The high- and low-temperature limits are also the same as Eqs.\ (\ref{F1})--(\ref{S1}) with $n=3$ and $\bar{E}^{(3)}=0$ (see Sec.\ \ref{Discussion}).

\begin{table*}
\caption{ \label{tab:C_F} Comparison of the zeroth-, through third-order corrections to the Helmholtz energy $F$ obtained from the $\lambda$-variation (numerical) method and 
sum-over-states (analytical) formulas as a function of temperature $T$ for an ideal gas of hydrogen fluoride in the canonical ensemble. }
\begin{ruledtabular}
\begin{tabular}{ldddddddd}
 &  \multicolumn{2}{c}{ $F^{(0)}/E_\text{h}$} &  \multicolumn{2}{c}{$ F^{(1)}/E_\text{h}$} &  \multicolumn{2}{c}{$ F^{(2)}/E_\text{h}$} &  \multicolumn{2}{c}{$ F^{(3)}/E_\text{h}$} \\ \cline{2-3}  \cline{4-5} \cline{6-7} \cline{8-9} 
$T /~\text{K}$ 
&\multicolumn{1}{c}{Numerical\tablenotemark[1]} &\multicolumn{1}{c}{Analytical\tablenotemark[2]}
&\multicolumn{1}{c}{Numerical\tablenotemark[1]} &\multicolumn{1}{c}{Analytical\tablenotemark[2]} 
&\multicolumn{1}{c}{Numerical\tablenotemark[1]} &\multicolumn{1}{c}{Analytical\tablenotemark[2]} 
&\multicolumn{1}{c}{Numerical\tablenotemark[1]} &\multicolumn{1}{c}{Analytical\tablenotemark[2]} 
\\  \hline
$10^4$ & -52.5749    & -52.5749    & -45.9959 & -45.9959 & -0.0173 & -0.0173 & -0.0055 & -0.0055 \\
$10^5$ & -52.6717    & -52.6717    & -46.1631 & -46.1631 & -0.1466 & -0.1466 & -0.0524 & -0.0524 \\
$10^6$ & -62.5554    & -62.5554    & -46.7786 & -46.7786 & -0.0165 & -0.0165 & 0.0003  & 0.0003  \\
$10^7$ & -176.803   & -176.803   & -46.8574 & -46.8574 & -0.0024 & -0.0024 & 0.0000  & 0.0000  \\
$10^8$ & -1368.93  & -1368.93  & -46.8576 & -46.8576 & -0.0004 & -0.0004 & 0.0000  & 0.0000  \\
$10^9$ & -13309.7 & -13309.7 & -46.8555 & -46.8555 & 0.0000  & 0.0000  & 0.0000  & 0.0000 
\end{tabular}
\tablenotetext[1]{The $\lambda$-variation benchmark, i.e., Eq.\ (\ref{eq:C_lambda_F}).}
\tablenotetext[2]{The sum-over-states analytical formula, i.e., Eq.\ (\ref{eq:zeroth_three}), (\ref{eq:first_three}), (\ref{eq:sec_three}), or (\ref{eq:three_three}). 
The first-, second-, and third-order HCPT energy corrections were evaluated by the $\lambda$-variation method as forward seven-point, seven-point, and five-point finite differences, respectively, with $\Delta\lambda = 10^{-3}$.}
\end{ruledtabular}
\end{table*}

\begin{table*}
\caption{ \label{tab:C_U} The same as Table \ref{tab:C_F} but for the internal energy $U$.}  
\begin{ruledtabular}
\begin{tabular}{ldddddddd}
 &  \multicolumn{2}{c}{ $U^{(0)}/E_\text{h}$} &  \multicolumn{2}{c}{$ U^{(1)}/E_\text{h}$} &  \multicolumn{2}{c}{$ U^{(2)}/E_\text{h}$} &  \multicolumn{2}{c}{$ U^{(3)}/E_\text{h}$} \\ \cline{2-3}  \cline{4-5} \cline{6-7} \cline{8-9} 
$T /~\text{K}$ 
&\multicolumn{1}{c}{Numerical\tablenotemark[1]} &\multicolumn{1}{c}{Analytical\tablenotemark[2]}
&\multicolumn{1}{c}{Numerical\tablenotemark[1]} &\multicolumn{1}{c}{Analytical\tablenotemark[2]} 
&\multicolumn{1}{c}{Numerical\tablenotemark[1]} &\multicolumn{1}{c}{Analytical\tablenotemark[2]} 
&\multicolumn{1}{c}{Numerical\tablenotemark[1]} &\multicolumn{1}{c}{Analytical\tablenotemark[2]} 
\\  \hline
$10^4$ & -52.5749 & -52.5749 & -45.9959 & -45.9959 & -0.0173 & -0.0173 & -0.0055 & -0.0055 \\
$10^5$ & -52.2645 & -52.2645 & -45.6944 & -45.6944 & -0.0215 & -0.0215 & -0.1665 & -0.1665 \\
$10^6$ & -50.6228 & -50.6228 & -46.7166 & -46.7166 & -0.0342 & -0.0342 & 0.0009  & 0.0009  \\
$10^7$ & -46.0028 & -46.0028 & -46.8452 & -46.8452 & -0.0037 & -0.0037 & 0.0001  & 0.0001  \\
$10^8$ & -42.4046 & -42.4046 & -46.8596 & -46.8596 & -0.0008 & -0.0008 & 0.0000  & 0.0000  \\
$10^9$ & -41.9496 & -41.9496 & -46.8557 & -46.8557 & -0.0001 & -0.0001 & 0.0000  & 0.0000 
\end{tabular}
\tablenotetext[1]{The $\lambda$-variation benchmark, i.e., Eq.\ (\ref{eq:C_lambda_F}).}
\tablenotetext[2]{The sum-over-states analytical formula, i.e., Eq.\ (\ref{eq:zeroth_four}), (\ref{eq:first_four}), (\ref{eq:sec_four}), or (\ref{eq:three_four}). See the 
corresponding caption of Table \ref{tab:C_F} for the evaluation of the HCPT energy corrections.}
\end{ruledtabular}
\end{table*}

\section{Discussion\label{Discussion}}

Table \ref{tab:HF_F} shows that the free energies in the canonical ($F$) and grand canonical ($\Omega$) ensembles \cite{jha} 
differ considerably from each other. At any temperature, the $n$th-order  ($0 \leq n \leq 2$) perturbation approximation to $\Omega$ (i.e., the sum of zeroth- through $n$th-order perturbation corrections) is 
always more negative than the corresponding perturbation approximation of $F$. 
The majority of the difference is accounted for by the $\mu \bar{N}$ contribution (where $\mu$ is the chemical potential 
and $\bar{N}$ is the average number of electrons canceling the positive nuclear charge).
While these two ensembles should be equivalent in the limit of large volume \cite{francisco}, 
with a minimal volume containing one molecule, the two sets of the results are far from convergence, 
suggesting that the canonical ensemble may not be used interchangeably with the grand canonical ensemble 
if there is any possibility of an electron hopping from one molecule to another, even though the charge neutrality of the system is always maintained in both ensembles \cite{jha}. 

Even in the zero-temperature limit, $F^{(n)}$ and $\Omega^{(n)}$ generally differ from each other except for $n=1$. This is because we can write the limits \cite{so_jha} as
\begin{eqnarray}
\lim_{T \to 0} F^{(n)}  &=& E_0^{(n)},\\
\lim_{T \to 0} \Omega^{(n)}  &=& E_0^{(n)} - \mu^{(n)} \bar{N},
\end{eqnarray}
where $E_0^{(n)}$ is the $n$th-order correction to energy according to MPPT 
in the case of a nondegenrate $N$-electron ground-state wave function or HCPT in the case of a degenerate $N$-electron ground-state wave function.
Only in the first order for a nondegenerate ground state, $\mu^{(1)}=0$ and, therefore, $F^{(1)} = \Omega^{(1)}$ at $T=0$ \cite{so_jha}.

The incorrect formulas for $\Omega^{(1)}$ and $\Omega^{(2)}$ (not shown) in various textbooks \cite{jha} give values that are 
more similar to $F^{(1)}$ and $F^{(2)}$ in the canonical ensemble, but are far from the correct values of $\Omega^{(1)}$ and $\Omega^{(2)}$ reproduced in Table \ref{tab:HF_F}. 
This is understandable because the grand-canonical 
theory in textbooks neglects to vary $\mu$, while the canonical ensemble does not have $\mu$ in the first place. However, it should be remembered that 
the canonical ensemble of a neutral system is valid thermodynamics,
whereas the grand canonical ensemble of a massively charged system is not.

Comparing the Helmholtz ($F$) and internal ($U$) energies in the canonical ensemble compiled in Tables \ref{tab:HF_F} and \ref{tab:HF_U}, 
we observe that $F^{(n)} = U^{(n)}$ at $T=0$ for $0 \leq n \leq 3$. In fact, since 
\begin{eqnarray}
F^{(n)} = U^{(n)} - TS^{(n)},
\end{eqnarray}
this is expected to hold true for any $n$. 

Table \ref{tab:HF_U} shows that the internal energies $U^{(n)}$ in the canonical and grand canonical  \cite{jha} ensembles converge at the same zero-temperature limit for any $n$.
This is again expected because 
\begin{eqnarray}
\lim_{T \to 0} U^{(n)}  &=& E_0^{(n)},
\end{eqnarray}
in both ensembles.

As the temperature increases, $U^{(n)}$ tends to two distinct limits depending on the ensembles. 
In the canonical ensemble, the high-temperature limit of $U^{(n)}$ is the average of $E_I^{(n)}$ over all $N$-electron states \cite{Kou2014}:
\begin{eqnarray}
\lim_{T \to \infty} U^{(n)} = \frac{\sum_I E_I^{(n)}}{{}_{2k}C_N}, \label{trace}
\end{eqnarray} 
where the denominator is the total number of $N$-electron states.
In the grand canonical ensemble, $U^{(n)}$ has a different limit \cite{Kou2014},
\begin{eqnarray}
\lim_{T \to \infty} U^{(n)} =\frac{\sum_I E_I^{(n)} \{{N} / (2k - {N})\}^{N_I}}{\sum_I \{{N} / (2k - {N})\}^{N_I}},
\end{eqnarray} 
where $I$ runs over all states whose electron count $N_I$ ranges from zero to $2k$ ($k$ is the number of basis functions) \cite{Kou2014}; it
is not a simple average of energies.
In either case, these limiting behaviors are often an artifact of a finite number of basis functions, and are dependent on $k$. 

In the canonical ensemble, we observe 
\begin{eqnarray}
\lim_{T \to \infty} U^{(n)} &=& 0\,\,\, \text{for~}n\geq 2.
\end{eqnarray}
This is explained by the similarity-invariance of trace. Equation (\ref{trace}) means that 
$U^{(0)} + U^{(1)}$ is the trace of the Hamiltonian in the complete $N$-electron determinant basis divided by ${}_{2k}C_N$, which is already exact in the finite basis set.
Therefore, $U^{(2)}$ and higher-order corrections are zero in this limit. We believe that this is not an artifact of a finite basis set; an ensemble average 
of energy has less correlation because of mutual cancellation of correlation energies among ground and excited states, which tends to null correlation in the high-temperature limit.

In a finite-basis theory \cite{Kou2014}, we have
\begin{eqnarray}
\lim_{T \to 0} S  &=& 0,\\
\lim_{T \to \infty} S &=& k_\text{B} \ln {}_{2k}C_{N}.
\end{eqnarray}
The former follows from Nernst's theorem and the latter (entropy saturation) is a finite-basis-set artifact.
These relations in conjunction with Eq.\ (\ref{eq:C_lambda_F}) imply
\begin{eqnarray}
\lim_{T \to 0} S^{(n)} &=& 0,\\
\lim_{T \to \infty} S^{(0)} &=& k_\text{B} \ln {}_{2k}C_{N}, \\
\lim_{T \to \infty} S^{(n)} &=& 0\,\,\, \text{for~}n\geq 1,
\end{eqnarray}
which are numerically verified in Table \ref{tab:HF_S}.

Entropy is always greater in the grand canonical ensemble than in the canonical ensemble at any temperature, which is intuitive, but 
the behavior of its perturbation corrections is hard to predict. 

Table \ref{tab:diff} lists the deviation of the sum of zeroth- through third-order corrections  
from the thermal-FCI value for $F$, $U$, and $S$ at various temperatures. In all cases, it 
shows  rapid convergence of the perturbation series. An exception occurs at $10^5$~K,
where the third-order perturbation theory has an error of $13\,\text{m}E_\text{h}$ for $F$, 
$31\,\text{m}E_\text{h}$ for $U$, and 5\% for $S$. The slow convergence coincides with the rapid rise
in $F$, $U$, and $S$ at around $10^5$~K, which roughly corresponds to the lowest excitation 
energy of the hydrogen fluoride molecule in the minimal basis set \cite{Kou2014}. 
Below this temperature, the convergence of $F$ and $U$ is essentially the same as 
that of zero-temperature MPPT (which also has an error of $3\,\text{m}E_\text{h}$ at the third order). 
Above this temperature, the convergence is extremely rapid, which may be interpreted to support the notion
that strong correlation can be more easily described at higher temperatures even by perturbation theory.
It may instead be an artifact of the smallness of 
the basis set used.  

Tables \ref{tab:C_F} and \ref{tab:C_U} underscore the numerically exact agreement between the sum-over-states
analytical formulas and the $\lambda$-variation benchmark data for $F^{(n)}$ and $U^{(n)}$ ($0 \leq n \leq 3$).
It mutually verifies the analytical formulas and the precision of the $\lambda$-variation calculations at all temperatures studied.

Similar observations can be made to the benchmark data of the perturbation corrections for the boron hydride and 
beryllium atom, which are recorded in the Appendix. 

\section{Conclusions}

We have documented the benchmark data for the zeroth- through third-order perturbation corrections
to the Helmholtz energy, internal energy, and entropy in the canonical ensemble 
for several ideal gases of atoms or molecules in a wide range of temperature.

We have also presented the sum-over-states analytical formulas for these perturbation corrections expressed in terms of HCPT energy corrections.
These benchmark data and analytical formulas have been mutually verified by exact numerical agreement. 
 We have not found a kind of mathematical reduction that compresses the sum-over-states formulas in the grand canonical ensemble 
 to the formulas expressed in terms of molecular integrals and the Fermi--Dirac distribution function \cite{so_jha}.
 
 The perturbation corrections to the internal energies are close to each other between the canonical ensemble and grand canonical ensemble, when both
 maintain the charge neutrality \cite{jha}; they may be used interchangeably. The perturbation corrections to the free energies (Helmholtz energy in the canonical ensemble
 and grand potential in the grand canonical ensemble) are, on the other hand, rather different because of the $\mu^{(n)}\bar{N}$ contribution in the latter.
 
 For these two reasons (the lack of mathematical reduction and the poor convergence to the grand canonical ensemble
 for the smallest volume), the utility of the canonical ensemble for electrons may be somewhat limited for computing the free energy.

\acknowledgments
This work was supported by the Center for Scalable, Predictive methods for Excitation and Correlated phenomena (SPEC), which is funded by the U.S. Department of Energy, Office of Science, Office of Basic Energy Sciences, Chemical Sciences, Geosciences, and Biosciences Division, as a part of the Computational Chemical Sciences Program and
also by the U.S. Department of Energy, Office of Science, Office of Basic Energy Sciences under Grant No.\ DE-SC0006028.

\appendix*

\section{$\lambda$-variation numerical benchmarks for BH and Be}

Tables \ref{tab:BH_F} through \ref{tab:BH_S} document the zeroth- through third-order perturbation corrections to 
the Helmholtz energy $F$, internal energy $U$, and entropy $S$, respectively, for an ideal gas of the boron hydride 
molecule (1.232~\AA) in the STO-3G basis set in the canonical ensemble. 
Tables \ref{tab:Be_F} through \ref{tab:Be_S} compile the same for an ideal gas of the beryllium atom in the STO-3G basis set in the canonical ensemble. 

These data, along with the one presented in the main text, are hoped to serve as a useful benchmark for testing or calibrating analytical formulas
or other approximations.

\begin{table} 
\caption{ \label{tab:BH_F} The zeroth- through third-order perturbation corrections to the Helmholtz energy $F$ as a function of temperature $T$ 
obtained from the $\lambda$-variation method for an ideal gas of boron hydride in the canonical ensemble. }
\begin{ruledtabular}
\begin{tabular}{ldddd}
$T /~\text{K}$ &  \multicolumn{1}{c}{$ F^{(0)}/E_\text{h}$ } &  \multicolumn{1}{c}{ $F^{(1)}/E_\text{h}$} & \multicolumn{1}{c}{$F^{(2)}/E_\text{h}$} & \multicolumn{1}{c}{$F^{(3)}/E_\text{h}$}  \\ \hline
$10^3$ & -14.1712    & -10.5816 & -0.0295 & -0.0134 \\
$10^4$ & -14.1712    & -10.5816 & -0.0295 & -0.0135 \\
$10^5$ & -14.6289    & -11.0154 & -0.1712 & -0.0166 \\
$10^6$ & -29.8911    & -11.6495 & -0.0370 & -0.0003 \\
$10^7$ & -221.425  & -11.7999 & -0.0082 & 0.0000  \\
$10^8$ & -2167.32  & -11.7767 & -0.0009 & 0.0000  \\
$10^9$ & -21629.9 & -11.7737 & -0.0001 & 0.0000 
\end{tabular}
\end{ruledtabular}
\end{table}

\begin{table}  
\caption{ \label{tab:BH_U} The same as Table \ref{tab:BH_F} but for the internal energy $U$. }
\begin{ruledtabular}
\begin{tabular}{ldddd}
$T /~\text{K}$ &  \multicolumn{1}{c}{$ U^{(0)}/E_\text{h}$} &  \multicolumn{1}{c}{$ U^{(1)}/E_\text{h}$} & \multicolumn{1}{c}{$U^{(2)}/E_\text{h}$} & \multicolumn{1}{c}{$U^{(3)}/E_\text{h}$}  \\ \hline
$10^3$ & -14.1712 & -10.5816 & -0.0295 & -0.0134 \\
$10^4$ & -14.1712 & -10.5816 & -0.0295 & -0.0133 \\
$10^5$ & -13.5208 & -10.5793 & -0.2592 & -0.0402 \\
$10^6$ & -10.8720 & -11.3909 & -0.0507 & -0.0013 \\
$10^7$ & -5.5759  & -11.8196 & -0.0156 & 0.0001  \\
$10^8$ & -4.8512  & -11.7799 & -0.0018 & 0.0000  \\
$10^9$ & -4.7785  & -11.7740 & -0.0002 & 0.0000 
\end{tabular}
\end{ruledtabular}
\end{table}

\begin{table} 
\caption{ \label{tab:BH_S} The same as Table \ref{tab:BH_F} but for the entropy $S$.}
\begin{ruledtabular}
\begin{tabular}{ldddd}
$T /~\text{K}$ &  \multicolumn{1}{c}{ $S^{(0)}/k_\text{B}$} &  \multicolumn{1}{c}{$ S^{(1)}/k_\text{B}$} & \multicolumn{1}{c}{$S^{(2)}/k_\text{B}$} & \multicolumn{1}{c}{$S^{(3)}/k_\text{B}$}  \\ \hline
$10^3$ & 0.0000 & 0.0000  & 0.0000  & 0.0000  \\
$10^4$ & 0.0000 & 0.0002  & 0.0011  & 0.0054  \\
$10^5$ & 3.4991 & 1.3772  & -0.2777 & -0.0745 \\
$10^6$ & 6.0058 & 0.0817  & -0.0043 & -0.0003 \\
$10^7$ & 6.8160 & -0.0006 & -0.0002 & 0.0000  \\
$10^8$ & 6.8286 & -0.0000 & 0.0000  & 0.0000  \\
$10^9$ & 6.8287 & -0.0000  & -0.0000  & 0.0000 
\end{tabular}
\end{ruledtabular}
\end{table}


\begin{table} 
\caption{ \label{tab:Be_F}  The zeroth- through third-order perturbation corrections to the Helmholtz energy $F$ as a function of temperature $T$ 
obtained from the $\lambda$-variation method for an ideal gas of beryllium in the canonical ensemble.  }
\begin{ruledtabular}
\begin{tabular}{ldddd}
$T /~\text{K}$ &  \multicolumn{1}{c}{$ F^{(0)}/E_\text{h}$ } &  \multicolumn{1}{c}{ $F^{(1)}/E_\text{h}$} & \multicolumn{1}{c}{$F^{(2)}/E_\text{h}$} & \multicolumn{1}{c}{$F^{(3)}/E_\text{h}$}  \\ \hline
$10^3$ & -9.4761     & -4.8758 & -0.0244 & -0.0140 \\
$10^4$ & -9.4761     & -4.8758 & -0.0244 & -0.0140 \\
$10^5$ & -9.9469     & -5.2087 & -0.0803 & 0.0065  \\
$10^6$ & -21.6451    & -5.5326 & -0.0238 & 0.0006  \\
$10^7$ & -172.736   & -5.4445 & -0.0048 & 0.0000  \\
$10^8$ & -1696.59  & -5.4192 & -0.0005 & 0.0000  \\
$10^9$ & -16936.4 & -5.4165 & -0.0001 & 0.0000 
\end{tabular}
\end{ruledtabular}
\end{table}

\begin{table}  
\caption{ \label{tab:Be_U} The same as Table \ref{tab:Be_F} but for the internal energy $U$. }
\begin{ruledtabular}
\begin{tabular}{ldddd}
$T /~\text{K}$ &  \multicolumn{1}{c}{$ U^{(0)}/E_\text{h}$} &  \multicolumn{1}{c}{$ U^{(1)}/E_\text{h}$} & \multicolumn{1}{c}{$U^{(2)}/E_\text{h}$} & \multicolumn{1}{c}{$U^{(3)}/E_\text{h}$}  \\ \hline
$10^3$ & -9.4761 & -4.8758 & -0.0244 & -0.0140 \\
$10^4$ & -9.4761 & -4.8758 & -0.0243 & -0.0136 \\
$10^5$ & -9.0282 & -5.0131 & -0.1728 & 0.0091  \\
$10^6$ & -6.1047 & -5.4802 & -0.0289 & 0.0003  \\
$10^7$ & -3.5488 & -5.4712 & -0.0093 & 0.0000  \\
$10^8$ & -3.2885 & -5.4221 & -0.0010 & 0.0000  \\
$10^9$ & -3.2627 & -5.4168 & -0.0001 & -0.0000
\end{tabular}
\end{ruledtabular}
\end{table}

\begin{table}   
\caption{ \label{tab:Be_S} The same as Table \ref{tab:Be_F} but for the entropy $S$. }
\begin{ruledtabular}
\begin{tabular}{ldddd}
$T /~\text{K}$ &  \multicolumn{1}{c}{ $S^{(0)}/k_\text{B}$} &  \multicolumn{1}{c}{$ S^{(1)}/k_\text{B}$} & \multicolumn{1}{c}{$S^{(2)}/k_\text{B}$} & \multicolumn{1}{c}{$S^{(3)}/k_\text{B}$}  \\ \hline
$10^3$ & 0.0000 & 0.0000  & 0.0000  & 0.0000  \\
$10^4$ & 0.0001 & 0.0006  & 0.0035  & 0.0132  \\
$10^5$ & 2.9011 & 0.6175  & -0.2922 & 0.0081  \\
$10^6$ & 4.9073 & 0.0166  & -0.0016 & -0.0001 \\
$10^7$ & 5.3425 & -0.0008 & -0.0001 & 0.0000  \\
$10^8$ & 5.3471 & -0.0000 & -0.0000  & 0.0000  \\
$10^9$ & 5.3471 & -0.0000  & -0.0000  & 0.0000 
\end{tabular}
\end{ruledtabular}
\end{table}

\bibliography{prl.bib}

\end{document}